\documentstyle[12pt,epsf,epsfig,a4]{article}
\begin{document}
\begin {titlepage}
\begin{flushleft}
FSUJ TPI QO-10a/98
\end{flushleft}
\begin{flushright}
July, 1998
\end{flushright}
\vspace{20mm}

\begin{center}
{\Large \bf Generation of arbitrary quantum states of 
traveling fields } 
\vspace{15mm}
 
{\large \bf M. Dakna, J. Clausen, L. Kn\"oll, and D.-G. Welsch}\\[1.ex]
{\large Friedrich-Schiller-Universit\"at Jena}\\[0.5 ex]
{\large Theoretisch-Physikalisches Institut}\\[0.5 ex]
{\large Max-Wien Platz 1, D-07743 Jena, Germany}
\vspace{25mm}
\end{center}
\begin{center}
\bf{Abstract}
\end{center}
We show that any single-mode quantum state can be generated
from the vacuum by alternate application of the
coherent displacement operator and the creation operator.
We propose an experimental implementation of the scheme
for traveling optical fields, which is based on field
mixings and conditional measurements in a beam splitter
array, and calculate the probability of state generation. 
\end{titlepage}

Designing of schemes for the generation of specific nonclassical 
quantum states has been a subject of increasing interest. The 
realization in the laboratory of schemes that have been already proposed  
has been one of the most exciting challenges to the researchers. 
In \cite{Vogel} a method is proposed that offers the possibility
of preparing a cavity-field mode undergoing a Jaynes-Cummings dynamics
in any superposition of a finite number of Fock states in principle.
The method is based on a non-unitary ``collapse'' of the state 
vector of the cavity-field mode via atom ground-state measurement.
Before entering the cavity and interacting with the
cavity mode in a controlled way, the atoms are prepared in a 
well-defined superposition of two (Rydberg-)states.
After leaving the cavity, the atoms enter a detector for
measuring their energies.  

In this article we propose a scheme for the preparation of 
a radiation-field mode in an arbitrary (finite) superposition
of Fock states, by performing alternately coherent quantum-state 
displacement and single-photon adding in a well-defined succession. 
The advantage of the scheme is that it not only applies
to cavity-field modes but also to traveling-field modes. 
To be more specific, we first recall that
coherent quantum-state displacement can be realized for both 
cavity-field modes (see, e.g., \cite{Alsing}) and traveling-field modes 
(see, e.g., \cite{Ban}). In the former case an external
(classical) oscillator is resonantly coupled through one of the
mirrors to the cavity-field mode. In the latter case
the coherent displacement can be achieved with an 
appropriately chosen beam splitter for
mixing the signal mode with a strong local oscillator.
With regard to cavity-field modes, single-photon adding can be 
realized by injecting excited atoms into a cavity and
detecting the ground state of the atoms, after leaving the
cavity. Adopting the Jaynes-Cummings model, it can be shown 
\cite{Agarwal} that if an atom after interaction with a cavity-field 
mode is detected in the ground state, then the state of the 
cavity-field mode is reduced, under certain conditions, to 
$\sim\hat{a}^\dagger|\Phi\rangle$, $|\Phi\rangle$ being the state 
of the cavity-field mode before the atom enters the cavity. With regard 
to traveling-field modes, the non-unitary ``collapse'' to a photon-added 
state can be realized by conditional output measurement on a beam 
splitter \cite{Dakna}. In particular, when a mode prepared in a state 
$|\Phi\rangle$ is mixed at the beam splitter with a 
single-photon Fock state \cite{Kwiat} and a zero-photon measurement 
is performed in one of the output channels of the beam splitter, then 
the quantum state of the mode in the other output channels ``collapses'' 
to $\sim\hat{Y}|\Phi\rangle$, with
\begin{eqnarray}
\label{1.0}
\hat{Y} =  R\,\hat{a}^\dagger T^{\hat{n}} 
\end{eqnarray}
($R$, reflectance; $T$ transmittance of the beam splitter).
 
Let us assume that the quantum state that is desired to 
be generated is a finite superposition of Fock states, 
\begin{equation}
|\Psi\rangle=\sum_{n=0}^N \psi_n\,|n\rangle.
\label{1.1}
\end{equation} 
Note that the expansion of any physical state in the Fock basis can 
always be approximated to any desired degree of accuracy by truncating 
it at $N$ if $N$ is suitably large. 
Recalling the definition of Fock states, 
Eq.~(\ref{1.1}) can be given by
\begin{equation}
|\Psi\rangle=\sum_{n=0}^N \frac{\psi_n}{\sqrt{n!}}\,
(\hat{a}^\dagger)^n \,|0\rangle ,
\label{1.2}
\end{equation}
which may be rewritten as
\begin{eqnarray}
|\Psi\rangle\sim(\hat{a}^\dagger\!-\!\beta_N^\ast)(\hat{a}^\dagger\!-\!
\beta_{N-1}^\ast)\cdots(\hat{a}^\dagger\!-\!\beta_{2}^\ast)
(\hat{a}^\dagger\!-\!\beta_1^\ast)\,|0\rangle .
\label{1.3}
\end{eqnarray}
Here, $\beta_1^\ast,\beta_2^\ast,\cdots,\beta_N^\ast$ are the 
$N$ (complex) roots of the characteristic polynomial
\begin{equation}
\sum_{n=0}^N  \frac{\psi_n}{\sqrt{n!}} \, (\beta^\ast)^n = 0.
\label{1.4}
\end{equation}
Using the relation 
\begin{equation}
\hat{a}^\dagger\!-\!
\beta^\ast\!=\!\hat{D}(\beta)\hat{a}^\dagger\hat{D}^\dagger(\beta),
\label{1.5}
\end{equation}
where $\hat{D}(\beta)$ $\!=$ 
$\!\exp(\beta\hat{a}^\dagger$ $\!-$ $\!\beta^\ast\hat{a})$
is the coherent displacement operator, from Eq.~(\ref{1.3})
we find that  
\begin{eqnarray}
\lefteqn{
|\Psi\rangle\sim\hat{D}(\beta_N)\hat{a}^\dagger
\hat{D}^\dagger(\beta_N)
}
\nonumber \\ && \hspace{2ex}\times\,
\hat{D}(\beta_{N-1})\hat{a}^\dagger\hat{D}^\dagger(\beta_{N-1})
\cdots
\hat{D}(\beta_{1})\hat{a}^\dagger
\hat{D}^\dagger(\beta_1)\,|0\rangle .
\label{1.6}
\end{eqnarray}
Hence, any quantum state of the form (\ref{1.1}) can be obtained
from the vacuum by a succession of alternate state displacement
and single-photon adding, the displacements being determined
by the roots of the characteristic polynomial (\ref{1.4}). 

\begin{figure}
\centering\epsfig{figure=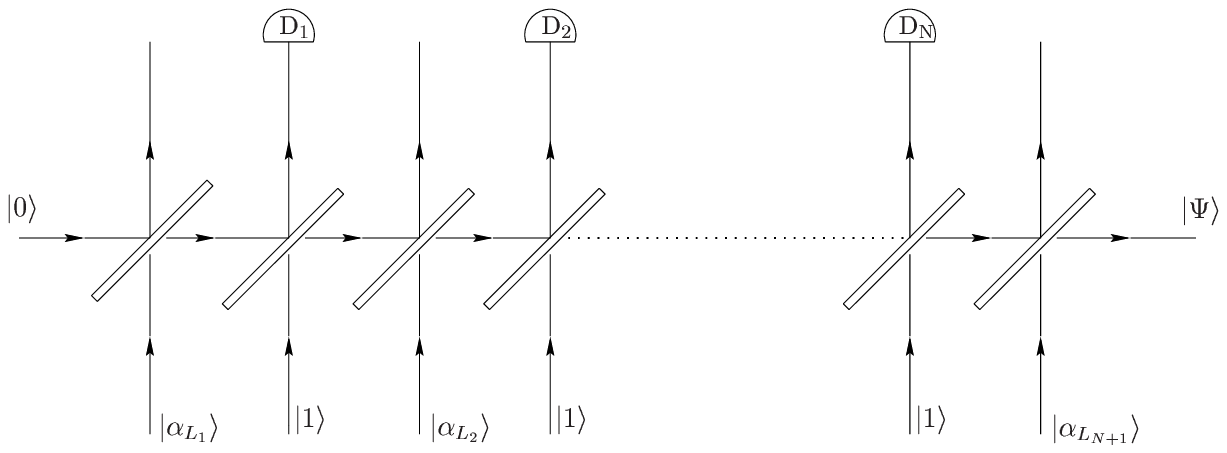,width=1\linewidth}
~

\vspace{4cm}
~
\caption{
Experimental setup for preparing a traveling-field mode in a  
quantum state $|\Psi\rangle$, Eq.~(\protect\ref{1.1}). 
At the first stage, a mode prepared in the vacuum state $|0\rangle$
and a mode prepared in a strong coherent state $|\alpha_{L_1}\rangle$
are superimposed by a beam splitter with transmittance
$\tilde{T}$ and reflectance $\tilde{R}$  
($|\tilde{T}|$ $\!\to$ $\!1$)
in order to produce a displaced vacuum state
$\hat{D}(\alpha_1)|0\rangle$ with $\alpha_1$ $\!=$ 
$\!\tilde{R}\alpha_{L_1}$. At the second stage, the mode 
prepared in the displaced vacuum state $\hat{D}(\alpha_1)|0\rangle$ 
and a mode prepared in a single-photon Fock state $|1\rangle$ are
superimposed by a beam splitter with transmittance $T$. When
the detector D$_1$ does not register photons, then the mode in the other
output channel of the beam splitter is prepared in a photon-added state
$\sim\hat{a}^\dagger\,T^{\hat{n}}\hat{D}(\alpha_1)\,|0\rangle$.
Now the two-step procedure is repeated, with the
states $\hat{a}^\dagger T^{\hat{n}}\hat{D}(\alpha_1)\,|0\rangle$
and $|\alpha_{L_2}\rangle$ in place of the states $|0\rangle$ and 
$|\alpha_{L_1}\rangle$, respectively. As a result, the state 
$\sim \hat{a}^\dagger T^{\hat{n}}\hat{D}(\alpha_2)
\hat{a}^\dagger T^{\hat{n}}\hat{D}(\alpha_1)\,|0\rangle$ is produced.
Repeating the procedure $N$ times and performing eventually an
additional state displacement $\hat{D}(\alpha_{N+1})$ obviously 
yields the state in Eq.~(\protect\ref{1.7}).
Choosing the values of $\alpha_{L_k}$ such that the values
$\alpha_{k}$ in Eqs.~(\protect\ref{1.11}) -- (\protect\ref{1.11b}) 
are realized, then the output state $|\Psi\rangle$ is
the desired state.   
\label{Fig1}
}
\end{figure}

An implementation of the method for a single-mode traveling field is 
outlined in Fig.~\ref{Fig1}. Following \cite{Dakna}, the state that is
produced if no photons are registered in each of the $N$ conditional
output measurements is given by
\begin{eqnarray}
\lefteqn{
|\Psi\rangle\sim\hat{D}(\alpha_{N+1})\hat{a}^\dagger T^{\hat{n}}
\hat{D}(\alpha_N)
}
\nonumber \\ && \hspace{6ex}\times\,
\hat{a}^\dagger T^{\hat{n}}\hat{D}(\alpha_{N-1})\cdots
\hat{a}^\dagger T^{\hat{n}}
\hat{D}(\alpha_1)\,|0\rangle .
\label{1.7}
\end{eqnarray}
In order to bring Eq.~(\ref{1.7}) into the form of Eq.~(\ref{1.6}), we
first write ($k$ $\!=$ $\!1,2,\ldots,N$)
\begin{equation}
T^{\hat{n}}\hat{D}(\alpha_k) = \hat{D}(\alpha_k)
\left[\hat{D}^\dagger(\alpha_k) T^{\hat{n}}\hat{D}(\alpha_k)\right]
\label{1.8}
\end{equation}
and then move the operators $\hat{D}^\dagger(\alpha_k) 
T^{\hat{n}}\hat{D}(\alpha_k)$ towards the right, on using the relation
\begin{eqnarray}
\lefteqn{
\left[\hat{D}^\dagger(\alpha)
 T^{\hat{n}}\hat{D}(\alpha)\right]\hat{a}^\dagger\!=
 \!T(\hat{a}^\dagger\!+\!\bar{T}\alpha^\ast)
\left[\hat{D}^\dagger(\alpha)
T^{\hat{n}}\hat{D}(\alpha)\right]
}
\nonumber \\ && \hspace{5ex} 
= T \hat{D}^\dagger(\bar{T}^\ast\alpha) \,\hat{a}^\dagger\,
\hat{D}(\bar{T}^\ast\alpha) \left[\hat{D}^\dagger(\alpha)
T^{\hat{n}}\hat{D}(\alpha)\right],
\label{1.9} 
\end{eqnarray}
where $\bar{T}$ $\!=$ $\!1$ $\!-$ $\!T^{-1}$. 
After some algebra we obtain
\begin{eqnarray}
\label{1.10}
\lefteqn{
|\Psi\rangle \sim 
\hat{D}(\alpha_{N+1})\,\hat{a}^\dagger \hat{D}(T^*{}^{-1}\alpha_N)
}
\nonumber\\&&\hspace{1ex}\times\,
\hat{a}^\dagger \hat{D}(T^*{}^{-2}\alpha_{N-1})
\cdots
\hat{a}^\dagger \hat{D}(T^*{}^{1-N}\alpha_2) 
\nonumber\\&&\hspace{1ex}\times\,
\hat{a}^\dagger
\hat{D}\!\left(T^N\alpha_1 + \sum_{l=1}^N
[1\!-\!|T|^{2(l-N)}]T^{N-l}\alpha_{l+1}\right)
|0\rangle .
\end{eqnarray}
Comparing Eqs.(\ref{1.6}) and (\ref{1.10}), we find that
the two equations become identical, if the experimental 
displacement parameters $\alpha_k$ are chosen as follows:
\begin{eqnarray}
\label{1.11}
&& \alpha_1= - \sum_{l=1}^{N}T^{-l}\alpha_{l+1}\,,
\\
\label{1.11a}
&& \hspace{0ex}
\alpha_k= T^{\ast}{^{N-k+1}}(\beta_{k-1}-\beta_k)\,,
\quad k=2,3,\ldots,N,
\\
\label{1.11b}
&& \hspace{0ex}
\alpha_{N+1}= \beta_N\,.
\end{eqnarray}
The numerical implementation of the method is rather simple. First
the roots of the polynomial (\ref{1.4}) are calculated, which can be
done using standard routines. A straightforward calculation then yields, 
according to Eqs.~(\ref{1.11})
-- (\ref{1.11b}), the displacement parameters required in the
experimental scheme.   

Let us address the question of what is the probability $P_{|\Psi\rangle}$
of producing a chosen state $|\Psi\rangle$.  
Obviously, this probability is determined by the requirement that
all the $N$ detectors do not register photons.
It can be given by 
\begin{eqnarray}
\lefteqn{
P_{|\Psi\rangle} 
= P(N,0|1,0;2,0;\ldots;N\!-\!1,0) \cdots
}
\nonumber \\ && \hspace{8ex}\times
\cdots P(2,0|1,0) \, P(1,0) .
\label{1.12}
\end{eqnarray}
Here, $P(k,0|1,0;2,0;\ldots;k\!-\!1,0)$ 
is the probability that the $k$th detector does not register photons
under the condition that the detectors D$_1$,D$_2$,\ldots,D$_{k-1}$ 
have also not registered photons.
To calculate the conditional probabilities in Eq.~(\ref{1.12}),
we note that the $k$th zero-photon measurement corresponds
to the application of the operator $\hat{Y}$, Eq.~(\ref{1.0}),
to the state resulting from the $(k\!-\!1)$th zero-photon measurement
(and subsequent displacement). Starting from
\begin{eqnarray}
\label{1.13a}
P(1,0) =\|\hat{Y}\hat{D}(\alpha_1)|0\rangle\|^2 ,
\end{eqnarray}
we derive step by step ($k$ $\!=$ $\!2,3,\ldots,N$) 
\begin{eqnarray}
\lefteqn{
P(k,0|1,0;2,0;\ldots;k\!-\!1,0)
}
\nonumber \\ && \hspace{2ex}=
\frac{\|\hat{Y}\hat{D}(\alpha_k)\hat{Y}\hat{D}(\alpha_{k-1})
\cdots\hat{Y}\hat{D}(\alpha_1) |0\rangle\|^2}
{\|\hat{Y}\hat{D}(\alpha_{k-1}) \cdots\hat{Y}\hat{D}(\alpha_1) |0\rangle\|
^2} 
\label{1.13}
\end{eqnarray}
($\|\,|\Phi\rangle\|$ $\!=$ $\!\sqrt{\langle\Phi|\Phi\rangle}$).
Combining Eqs.~(\ref{1.12}) and (\ref{1.13}), we find that
\begin{eqnarray}
\label{1.19}
P_{|\Psi\rangle}
= {\cal P}_N^2 ,
\end{eqnarray}
where 
\begin{eqnarray}
\label{1.19a}
{\cal P}_k & = & 
\|\hat{Y}\hat{D}(\alpha_k)\hat{Y}\hat{D}(\alpha_{k-1}) 
\cdots\hat{Y}\hat{D}(\alpha_1) |0\rangle\| .
\end{eqnarray}
Substituting in Eq.~(\ref{1.19a}) for $\hat{Y}$ Eq.~(\ref{1.0})
and using Eqs.~(\ref{1.8}), (\ref{1.9}), and (\ref{1.5}), after
some algebra we obtain
\begin{eqnarray}
\label{1.15}
\lefteqn{
{\cal P}_k^2
=|R|^{2k}|T|^{k(k-1)}
\bigg\|\prod_{m=1}^{k}\left(\hat{a}^\dagger\!+\!b_{mk}^\ast\right)
|\gamma_k\rangle \bigg\|^2
}
\nonumber \\ && \hspace{6ex} \times \,
\exp\!\left(-|R|^2
\sum_{m=1}^{k}\bigg|\sum_{j=1}^{m}T^{m-j}\alpha_j\bigg|^2\right)
\end{eqnarray}
where the abbreviations $b_{1k}$ $\!=$ $\!0$, 
\begin{eqnarray}
\label{1.16}
b_{mk} = -\sum_{j=0}^{m-2}T^{\ast}{^{-j-1}}\alpha_{k-j}\,,
\quad m = 2,3,\ldots,k,
\end{eqnarray}
and
\begin{eqnarray}
\label{1.16a}
\gamma_k =
\sum_{j=1}^{k}T^{k+1-j}\alpha_j 
\end{eqnarray}
have been introduced. To calculate the square of the norm of the  state 
in Eq.~(\ref{1.15}), we may write  
\begin{eqnarray}
\label{1.17}
\lefteqn{
\bigg\|\prod_{m=1}^{k}(\hat{a}^\dagger\!+\!b_{mk}^\ast)
|\gamma_k\rangle\bigg\|^2
}
\nonumber \\ && \hspace{1ex}
= \langle\gamma_k| \prod_{m=1}^{k}(\hat{a}\!+\!b_{mk})
\prod_{l=1}^{k}(\hat{a}^\dagger\!+\!b_{lk}^\ast) |\gamma_k\rangle
\nonumber \\ && \hspace{1ex}
= \sum_{m,l=0}^{k}
\bigg\{\bigg[
\sum_{i_1,\cdots,i_m}\hspace{-2ex}{^{^{^<}}}
b_{i_1k}\cdots b_{i_mk}\bigg]
\bigg[
\sum_{i_1,\cdots,i_l}\hspace{-2ex}{^{^{^<}}}
b_{i_1k}^\ast\cdots b_{i_lk}^\ast\bigg]
\nonumber \\ && \hspace{20ex} \times \,
\langle\gamma_k|\hat{a}^{k-m}\hat{a}^\dagger{^{k-l}}|\gamma_k\rangle
\bigg\},
\end{eqnarray}
where the symbol $\sum_{i_1,i_2,\cdots}\hspace{-6.5ex}{^{^{^<}}}\hspace{5ex}$ 
is used to indicate that the summation requires the condition 
$i_1$ $\!<$ $\!i_2$ $\!<$ $\!\cdots$ to be satisfied, and
\begin{eqnarray}
\label{1.18}
\lefteqn{
\langle\gamma_k|\hat{a}^{k-m}\hat{a}^\dagger{^{k-l}}|\gamma_k\rangle
}
\nonumber \\[1ex] && \hspace{2ex}
= \left\{
\begin{array}{ll}(k\!-\!m)!\gamma_k^\ast{^{m-l}}
{\rm L}_{k-m}^{m-l}\big(-|\gamma_k|^2\big)
& \ \mbox{if}\quad l<m,\\[2ex]
(k\!-\!l)!\gamma_k^{l-m}
{\rm L}_{k-l}^{l-m}\big(-|\gamma_k|^2\big)
& \ \mbox{if}\quad l\ge m
\end{array}\right.
\end{eqnarray}
[L$_n^m(x)$ being the generalized Laguerre polynomial]. 

\begin{figure}
\centering\epsfig{figure=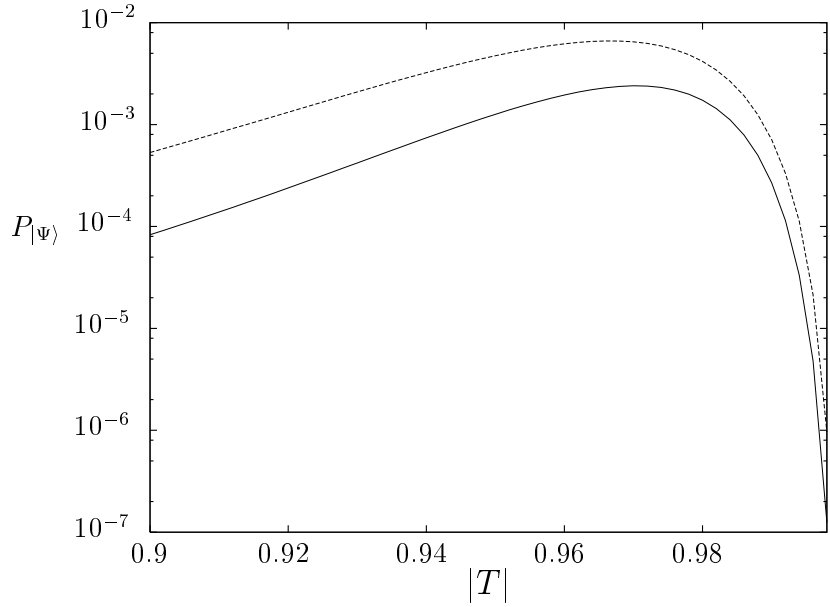,width=0.8\linewidth}
~

\vspace{4cm}
~
\caption{
The probability $P_{|\Psi\rangle}$ of producing truncated coherent phase
states $|\Psi\rangle$ $\!=$ $|z;N\rangle$, Eq.~(\protect\ref{1.20}), is 
shown as a function of the absolute value of the beam-splitter transmittance
$|T|$ for $N$ $\!=$ $\!6$ (solid line)
and $N$ $\!=$ $\!5$ (dotted line), and $z$ $\!=$ $\!0.4$.
It is assumed that the beam splitters used for photon adding
have the same transmittance.
\label{Fig2}
}
\end{figure}

In order to illustrate the method, let us consider the 
generation of truncated coherent phase states \cite{Shapiro}
\begin{eqnarray}
\label{1.20}
|\Psi\rangle \equiv
|z;N\rangle = C(z;N) \sum_{n=0}^{N}z^n\,|n\rangle ,
\end{eqnarray}
where
\begin{eqnarray}
\label{1.20b}
C(z;N) =
\left\{
\begin{array}{ll}
\displaystyle\sqrt{\frac{1\!-\!|z|^2}{1\!-\!|z|^{2(N+1)}}}
&\ {\rm if}\ |z|<1,
\\[1ex]
\displaystyle\frac{1}{\sqrt{N+1}}
&\ {\rm if}\ |z|=1.
\end{array}
\right.
\end{eqnarray}
The roots $\beta_k^\ast$ of the characteristic polynomial
\begin{eqnarray}
\label{1.20a}
\sum_{n=0}^N \frac{z^n}{\sqrt{n!}} \, (\beta^\ast)^n = 0
\end{eqnarray}
are given in Tab.~\ref{Tab1} for $z$ $\!=$ $\!0.4$ and $N$ $\!=$ $\!6$.
The table also shows the values of the displacement parameters 
$\alpha_k$ calculated from Eqs.~(\ref{1.11}) -- (\ref{1.11b}) 
for $T$ $\!=$ $\!0.99$. The probability of producing the state 
is $P_{|\Psi\rangle}$ $\!=$ $\!0.02\%$. For chosen state the 
probability $P_{|\Psi\rangle}$ sensitively depends on the absolute
value of the transmittance of the beam splitter, $|T|$,
as it can be seen from Fig.~\ref{Fig2} for two truncated 
coherent phase states. The probability increases with $|T|$, 
attains a maximum, and then rapidly approaches zero as $|T|$ 
goes to unity. So far we have assumed that the beam splitters used for
photon adding have the same transmittance. Assuming different
beam splitters, one may ask for the optimum set of transmittances 
that gives the highest probability of producing a chosen
state. Our numerical calculations for the truncated coherent
phase states has not led to a substantial improvement compared
to the case when equal beam splitters are used.
 
In summary, we have shown that single-mode radiation
can be prepared in arbitrary pure quantum states,
by a succession of alternate state displacement and single-photon 
adding. With regard to traveling fields, these operations
can be realized within a beam-splitter array into which 
coherent states and single-photon Fock states are fed
and zero-photon measurements are performed using
highly efficient avalanche photodiodes. It is worth noting that
the generation of arbitrary pure quantum states of traveling fields
offers new possibilities of quantum-state measurement, such
as projection synthesis for measuring the overlaps $|\langle A|\Phi\rangle|^2$
of a given state $|\Phi\rangle$ with arbitrary states $|A\rangle$
\cite{Barnett,Baseia}. Projection synthesis simply uses a beam splitter 
for combining the signal mode prepared in the state $|\Phi\rangle$ and a 
reference mode prepared in a state $|\Psi\rangle$ and two photodetectors 
in the output channels of the beam splitter for measuring the joint-event
probability distribution. The states $|\Psi\rangle$ can be calculated
from the states $|A\rangle$ (e.g., the states $|\Psi\rangle$ that are 
associated with the truncated phase states are reciprocal binomial
states \cite{Barnett}). Obviously, the crucial point is the preparation 
of specific states $|\Psi\rangle$, which may be solved
using the method proposed here. 

{\em Note added.} After preparing the article we were made aware
of a paper on the preparation of a superposition of the vacuum
and one-photon states of traveling fields by using similar
basic elements \cite{Barnett-Pegg-1998}.
 
\section*{Acknowledgements}
This work was supported by the Deutsche For\-schungs\-gemein\-schaft.

\bibliographystyle{unsrt}

\newpage
\newpage
\begin{table}
\centering\epsfig{figure=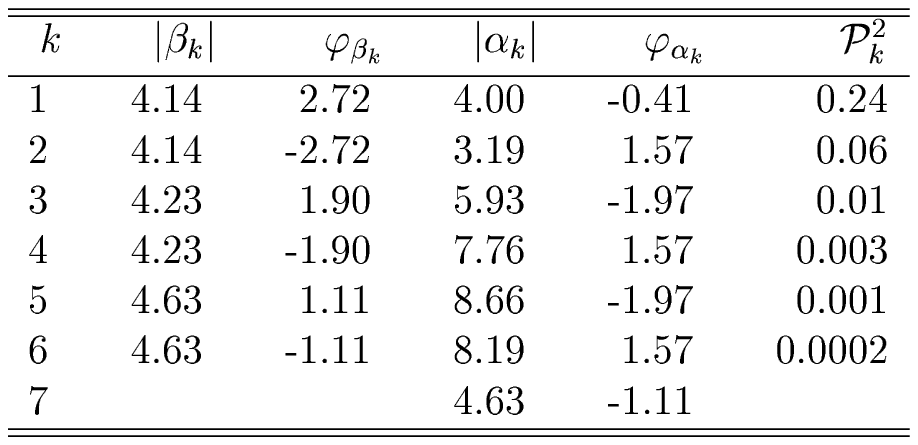,width=0.8\linewidth}
~

\vspace{5cm}
~
\caption{
The roots $\beta_k^\ast$ $\!=$ $|\beta_k|e^{-i\varphi_{\beta_k}}$
of the characteristic polynomial (\protect\ref{1.20a}) and the
displacement parameters $\alpha_k$ $\!=$ $|\alpha_k|e^{i\varphi_{\alpha_k}}$,
Eqs.~(\protect\ref{1.11}) -- (\protect\ref{1.11b}),
are given for a truncated coherent phase state $|\Psi$ $\!=$ $|z;N\rangle$
($z$ $\!=$ $\!0.4$, $N$ $\!=$ $\!6$), and $T$ $\!=$ $\!0.99$. 
The probability of producing the state is $P_{|\Psi\rangle}$
$\!=$ $\!0.02\%$. It is calculated from Eq.~(\protect\ref{1.19}),
the values of the ${\cal P}_k^2$ are given in the last column. 
\label{Tab1}
}
\end{table}
\newpage
\end{document}